\documentclass{article}
\usepackage[main, preprint]{neurips_2026}

\usepackage{microtype}
\usepackage{url}
\usepackage{booktabs}
\usepackage{algorithm}
\usepackage{algorithmic}
\usepackage{graphicx}
\usepackage{amsmath,amssymb}
\usepackage{xcolor}
\usepackage{multirow}
\usepackage{natbib}
\usepackage{enumitem}
\usepackage{hyperref}

\definecolor{darkblue}{rgb}{0, 0, 0.5}
\hypersetup{colorlinks=true, citecolor=darkblue, linkcolor=darkblue, urlcolor=darkblue}

\newcommand{\methodsc}{\textsc{SemanticVote}}

\title{Semantic Voting: Execution-Grounded Consensus for LLM Code Generation}

\author{
  Shan Jiang\thanks{Equal contribution, ordered by last name.} \quad
  Zijian Yi\footnotemark[1] \quad
  Chenguang Zhu \\
  The University of Texas at Austin \\
  \texttt{shanjiang@utexas.edu}
}

\begin{document}

\maketitle

\begin{abstract}
LLM code-generation pipelines often sample multiple candidate programs and then select one final answer without access to a complete oracle. Existing pipelines mix textual voting, ranking, and execution-based agreement, and the relative contribution of each component remains unclear. We study 18 configurations across different models, thinking levels, and benchmarks, comparing output-pattern majority voting, weighted voting, MBR-Exec, and \methodsc{}---a method that clusters candidates by execution fingerprints on LLM-generated inputs. Three findings emerge. (1)~The best execution-based selector exceeds output-pattern majority voting by 19--52 percentage points on every configuration, and every execution-based selector exceeds it by at least 18 points.
(2)~Once candidates are executed on diverse inputs, the aggregation rule has limited effect: \methodsc{}, weighted voting, and MBR-Exec are statistically indistinguishable across all 18 configurations. Paired bootstrap tests find no significant difference between \methodsc{} and either alternative ($p > 0.05$ throughout); \methodsc{} differs from weighted voting by only $-0.79$ to $+0.61$ pp. The largest observed factor is input quality: sketch-based input generation, in which an LLM enumerates abstract input categories before instantiating concrete values, is consistently strongest in the input-strategy ablation, exceeding direct LLM input generation by 0.6--2.1 pp and random fuzzing or example-only inputs by up to 11.3 pp; this benefit transfers consistently to all three execution-based aggregators. (3)~Thinking level interacts differently with the two selection families: on HumanEval+ Flash, deeper thinking improves majority voting by 12 pp, but execution-based methods stay flat or slightly degrade as candidate diversity falls. These results frame inference-time code selection as a signal-quality problem rather than an aggregation-rule problem: when oracles are unavailable, the behavioral evidence on which selection is based matters more than the rule used to aggregate it.
\end{abstract}

\section{Introduction}
\label{sec:intro}

Large language models (LLMs) are capable code generators, but selecting among multiple sampled candidate solutions remains difficult when no complete oracle is available. Existing systems combine textual voting, ranking heuristics, execution-based agreement, and learned reward models~\citep{wang2023selfconsistency,chen2023codet,shi2022mbr,li2022alphacode}, and the relative contribution of each component is unclear. We isolate two axes that prior work often conflates: (a) the value of richer behavioral evidence relative to all-or-nothing output-pattern voting, and (b) the relative importance of input quality and aggregation rule once selection uses behavioral evidence.

A common baseline ported from chain-of-thought reasoning is output-pattern majority voting (MV)~\citep{wang2023selfconsistency}: execute candidates on shared test inputs, discard any candidate that crashes on any input, group the remaining candidates by exact equality of concatenated return-value strings, and pick the largest group. This all-or-nothing rule is sensitive to generated-input errors because generated inputs are imperfect behavioral probes. A candidate that is correct on the benchmark may still raise on a generated input outside the prompt's intended domain, and majority voting then gives that candidate no vote. Conversely, if inputs miss a boundary case, an incorrect plurality can look identical to the correct behavior. The comparison therefore centers on how much behavioral evidence the selector preserves, rather than on execution alone.

We use \methodsc{} to study this comparison. \methodsc{} retains exception-aware execution fingerprints instead of reducing selection to an all-success output-pattern key. Given $N$ candidates, we execute each on $D$ test inputs and record an \emph{execution fingerprint}: the vector of outputs (or exception types) for each input. Programs with identical fingerprints form a semantic cluster, and the largest cluster's representative is returned. \methodsc{} sits in a family of execution-based methods alongside weighted voting~\citep{wang2025rankedvoting} and MBR-Exec~\citep{shi2022mbr}; we study the family rather than advocating for one member.

A precondition for any execution-based method is a set of test inputs diverse enough to distinguish semantically different programs. We introduce \emph{sketch-based input generation}: an LLM produces $K$ abstract input sketches (e.g., ``an empty list,'' ``a sorted list with duplicates,'' ``a list where all elements are negative''), each targeting a distinct behavioral equivalence class, and each sketch is instantiated $M$ times with concrete values, yielding $D = K \times M$ inputs.

\paragraph{Contributions.}
This is a findings paper. Across 18 configurations (3 Gemini models $\times$ 3 thinking levels $\times$ 2 benchmarks) we report three results:
\begin{itemize}[noitemsep,topsep=2pt]
    \item \textbf{Behavior over all-or-nothing voting.} The best execution-based selector outperforms output-pattern majority voting by 19--52 pp on every configuration, and output-pattern MV consistently scores below Best-of-$N$.
    \item \textbf{Input quality over aggregation rule.} Sketch-based input generation is consistently the strongest input strategy, exceeding direct LLM input generation by 0.6--2.1 pp and random fuzzing or example-only inputs by up to 11.3 pp. This gain transfers consistently to \methodsc{}, weighted voting, and MBR-Exec (at most 1.4 pp within strategy in the ablation). Once inputs are good, the three aggregators are statistically indistinguishable across all 18 configurations. Paired bootstrap tests find no significant difference between \methodsc{} and either alternative ($p > 0.05$ throughout); \methodsc{} differs from weighted voting by only $-0.79$ to $+0.61$ pp.
    \item \textbf{Different interactions with thinking depth.} On HumanEval+ Flash, deeper thinking improves output-pattern MV (+12 pp), while execution-based methods stay flat or slightly decrease as candidate diversity falls.
\end{itemize}
These results identify behavioral signal quality as the main factor in inference-time code selection.

\paragraph{Scope of contribution.}
The paper is an empirical decomposition rather than a method paper. Bootstrap analysis shows \methodsc{} $\approx$ weighted voting $\approx$ MBR-Exec at $p > 0.05$ in all 18 configurations. The supported conclusions are that execution-based selection outperforms all-or-nothing output-pattern voting, input quality drives most remaining variance, and thinking depth can affect the two selection families differently.

\section{Background and motivation}
\label{sec:background}

\paragraph{Inference-time scaling for code.}
Sampling more candidates and applying better selection improves code generation accuracy beyond greedy decoding \citep{chen2021codex,li2022alphacode,snell2025scaling}. This work studies candidate selection in settings where no complete oracle is available.

\paragraph{Output-pattern majority voting.}
Self-consistency \citep{wang2023selfconsistency} samples multiple chain-of-thought traces and picks the most common final answer. A common adaptation to code executes candidates on shared test inputs, discards candidates with any generated-input error, and clusters the survivors by exact equality of concatenated output strings. We treat this as a representative all-or-nothing output-pattern baseline: it uses execution, but it retains less behavioral evidence than exception-aware fingerprinting.

\paragraph{Execution-based selection.}
CodeT \citep{chen2023codet} generates test cases alongside code and ranks programs by dual execution agreement. \citet{shi2022mbr} apply minimum Bayes risk with execution on provided examples. AlphaCode \citep{li2022alphacode} clusters candidates by outputs on example tests. Our work differs in two ways: (a) sketch-based LLM-generated inputs in place of provided examples, and (b) exact-equality clustering on full execution fingerprints rather than pairwise agreement or output on a few examples.

\section{Method: \methodsc{}}
\label{sec:method}

\subsection{Pipeline overview}

\methodsc{} uses the LLM in exactly two places: generating candidate solutions (standard) and generating diverse test inputs via sketch-based input generation. The remaining steps---filtering, fingerprinting, clustering, and selection---are purely deterministic computation requiring no LLM calls. This separation is by design: it lets us isolate the effect of each component (input generation strategy, number of inputs, clustering method) in the ablation studies that follow, and it makes \methodsc{} a drop-in replacement for output-pattern voting at the same generation cost. Given a problem $P$ and an LLM $\mathcal{M}$, the pipeline proceeds in six steps (Algorithm~\ref{alg:pipeline}).

\begin{algorithm}[t]
\caption{\methodsc{} Pipeline}
\label{alg:pipeline}
\begin{algorithmic}[1]
\REQUIRE Problem $P$, model $\mathcal{M}$, budget $N$, sketches $K$, instantiations $M$
\ENSURE Selected solution $s^*$
\STATE $\{s_1, \ldots, s_N\} \leftarrow \text{Sample}(\mathcal{M}, P, N)$ \hfill \textit{// Generate candidates}
\STATE $S \leftarrow \text{Filter}(\{s_i\})$ \hfill \textit{// Remove syntactically invalid candidates}
\STATE $\{x_1, \ldots, x_D\} \leftarrow \text{SketchInputs}(\mathcal{M}, P, K, M)$ \hfill \textit{// Generate test inputs}
\FOR{each $s_i \in S$}
    \STATE $f_i \leftarrow [s_i(x_1), \ldots, s_i(x_D)]$ \hfill \textit{// Compute fingerprint}
\ENDFOR
\STATE $\mathcal{C} \leftarrow \text{ClusterByFingerprint}(\{(s_i, f_i)\})$ \hfill \textit{// Group by equality}
\STATE $C^* \leftarrow$ largest all-success cluster, else largest cluster \hfill \textit{// Prefer valid behavior}
\STATE $s^* \leftarrow \arg\min_{s \in C^*} |s|$ \hfill \textit{// Shortest program (Occam's razor)}
\RETURN $s^*$
\end{algorithmic}
\end{algorithm}

\subsection{Sketch-based input generation}
\label{sec:sketch-inputs}

We prompt the LLM to produce $K$ \emph{input sketches}: abstract descriptions of input categories targeting distinct behavioral equivalence classes. For a list-valued argument, example sketches include ``an empty list,'' ``a sorted list with duplicates,'' and ``a list with all negative numbers.'' Each sketch is instantiated $M$ times with concrete values, yielding $D = K \times M$ inputs. The two-level structure (categories first, instances second) gives diversity across input types and redundancy within each type.

\subsection{Execution fingerprinting}
\label{sec:fingerprint}

For each candidate $s_i$ and input $x_j$, we execute $s_i(x_j)$ in a sandboxed subprocess with a 5-second timeout and record one of: $(\texttt{ok}, \texttt{repr}(\text{result}))$ with floats rounded to 6 decimal places; $(\texttt{err}, \text{exception\_type})$ capturing the exception class name (not the message, which may be nondeterministic); or $(\texttt{timeout}, \texttt{TimeoutError})$. The fingerprint $f_i = [s_i(x_1), \ldots, s_i(x_D)]$ is a tuple of such pairs.

\subsection{Cluster selection}
\label{sec:selection}

Two candidates are placed in the same cluster iff $f_i = f_j$ (exact equality on all $D$ inputs). We return the shortest program (Occam's razor) from the largest cluster among those where all executions succeeded; if no all-success cluster exists, we fall back to the largest cluster overall.

\paragraph{Difference from output-pattern majority voting.}
Output-pattern MV, as adapted to code in prior work, also executes candidates, but it uses only all-success candidates and collapses each candidate to one concatenated output key. \methodsc{} instead keys on a tuple of $(\textit{status}, \texttt{repr}(\textit{value}))$ pairs, with floats rounded and container values serialized by Python after execution. This representation still distinguishes distinct values and exception types, but it does not discard a candidate's entire behavioral trace after a single generated-input error. At selection time, \methodsc{} prefers all-success clusters when they exist; if none exist, it falls back to the largest exception-aware fingerprint cluster. Weighted voting and MBR-Exec use the same execution traces in different ways, which lets us compare aggregation rules while holding the behavioral signal fixed.

\section{Experimental setup}
\label{sec:experiments}

\paragraph{Benchmarks.} We use HumanEval+ (164 problems) and MBPP+ (378 problems), both from EvalPlus \citep{liu2023evalplus}. EvalPlus augments HumanEval with roughly 80$\times$ more tests and MBPP with roughly 35$\times$ more tests, making pass@1 evaluation more reliable than the original small test suites. MBPP+ is built from the hand-verified MBPP-sanitized subset rather than the full MBPP collection; the current EvalPlus release contains 378 tasks after removing broken or ill-formed tasks. We use the unmodified EvalPlus releases of both benchmarks.

\paragraph{Models and thinking levels.} We evaluate three preview Gemini endpoints, using the exact model IDs accepted by the API in our runs: \texttt{gemini-3.1-pro-preview}, \texttt{gemini-3-flash-preview}, and \texttt{gemini-3.1-flash-lite-preview}. Tables abbreviate these as \textbf{3.1 Pro}, \textbf{3 Flash}, and \textbf{3.1 Flash Lite}. Each endpoint is evaluated at three \texttt{thinking\_level} settings (\texttt{low}, \texttt{medium}, \texttt{high}), which control the budget allocated to internal reasoning before output. The resulting $3\times 3$ grid lets us vary model capability and inference-time thinking level independently, while staying within a single model family to control for architecture and training-data differences.

\paragraph{Baselines.} We compare against the following baselines:
\begin{itemize}[noitemsep,topsep=2pt]
    \item \textbf{Greedy}: temperature 0, single sample.
    \item \textbf{Best-of-$N$}: first syntactically valid candidate.
    \item \textbf{Majority voting}: execute all candidates, discard any candidate with a generated-input error, group by exact concatenated output-pattern equality, return the shortest program from the largest group.
    \item \textbf{AST-normalized MV}: parse each candidate into an AST, alpha-rename variables, strip docstrings, and group by canonical AST structure---a purely code-structural baseline that correctly groups candidates differing only in naming or formatting, but cannot distinguish semantically different programs with identical structure.
    \item \textbf{Weighted voting}: output-pattern grouping with weights equal to per-candidate execution success rate.
    \item \textbf{MBR-Exec} \citep{shi2022mbr}: per-candidate score is the sum, over inputs, of how many other candidates agree on that input's output or exception type; pick the highest-scoring.
\end{itemize}
All non-greedy baselines share the same $N$ cached candidates and the same test inputs as \methodsc{}.

\paragraph{Configuration.} Unless stated otherwise, we sample $N = 50$ candidates at temperature 0.8 and remove syntactically invalid generations before selection. Sketch-based inputs use $K = 10$ and $M = 5$ ($D = 50$). Execution timeout is 5\,s per (candidate, input). The input-strategy ablation in Section~\ref{sec:ablation-strategy} uses cached $N = 100$ candidate pools held fixed across the four input strategies and three execution-based aggregators.

\paragraph{Greedy is a per-model baseline.} Greedy generations were cached per (model, prompt) without a thinking-level dimension before the cache key was extended to include thinking, so within a model the same single greedy sample populates the Greedy cells of every thinking-level row. The HumanEval+ Best-of-$N$ column inherits the same cached generations and is also per-model on that benchmark; on MBPP+, the candidate pools were resampled later and the Best-of-$N$ column does vary per thinking level. The three voting aggregators (MV, WV, MBR-Exec) and \methodsc{} are re-sampled per thinking level and reflect distinct candidate pools per cell. None of the paper's claims compare Greedy across thinking levels; Greedy and the HumanEval+ Best-of-$N$ row are used only as per-model floors against which voting methods are measured.

\section{Results}
\label{sec:results}

We organize the results around the two axes introduced above. Section~\ref{sec:ablation-strategy} addresses input quality first, since it is the largest observed source of variance among execution-based methods. Sections~\ref{sec:main-results}--\ref{sec:mbpp-results} then compare aggregation rules at fixed input quality. Section~\ref{sec:oracle-gap} decomposes failures into generation and selection components.

\subsection{Input generation strategy is the primary lever}
\label{sec:ablation-strategy}

A precondition for any execution-based method is a set of test inputs diverse enough to distinguish semantically different programs. We hold the candidate pool fixed and vary the input source across four strategies, evaluating each under all three execution-based selection methods:
\begin{itemize}[noitemsep,topsep=2pt]
    \item \textbf{Sketch (ours)}: $K$ abstract categories, each instantiated $M$ times (Section~\ref{sec:sketch-inputs}).
    \item \textbf{Direct LLM}: $D$ concrete inputs generated in one prompt, without the sketch abstraction.
    \item \textbf{Random}: type-aware random fuzzing.
    \item \textbf{Example-only}: only the example inputs from the problem description (typically 1--3).
\end{itemize}

\begin{table}[H]
\begin{center}
\small
\begin{tabular}{l ccc ccc ccc}
\toprule
& \multicolumn{3}{c}{\textbf{Pro low} (139)} & \multicolumn{3}{c}{\textbf{Flash med} (142)} & \multicolumn{3}{c}{\textbf{Lite low} (164)} \\
\cmidrule(lr){2-4} \cmidrule(lr){5-7} \cmidrule(lr){8-10}
\textbf{Strategy} & SV & WV & MBR & SV & WV & MBR & SV & WV & MBR \\
\midrule
Sketch (ours) & 99.3 & \textbf{100.0} & 98.6 & \textbf{97.2} & \textbf{97.9} & 96.5 & \textbf{95.1} & \textbf{95.1} & 94.5 \\
Direct LLM    & 97.8 & 98.6 & 97.8 & 95.8 & 95.8 & 94.4 & 94.5 & 94.5 & 93.9 \\
Random        & 94.2 & 94.2 & 95.0 & 89.4 & 89.4 & 88.7 & 93.9 & 93.9 & 93.9 \\
Example-only  & 95.0 & 95.0 & 94.2 & 86.6 & 86.6 & 85.9 & 90.9 & 90.9 & 90.9 \\
\bottomrule
\end{tabular}
\end{center}
\caption{Pass@1 (\%) on HumanEval+ across four input generation strategies and three execution-based aggregators, using cached $N=100$ candidate pools held fixed within each configuration. Problem counts (in parentheses) restrict to problems where input generation succeeded for all strategies. Within each row, the three aggregators agree to within 1.4 pp; within a fixed model and aggregator, the strategy choice changes accuracy by up to 11.3 pp.}
\label{tab:ablation-strategy}
\end{table}

On Flash medium, sketch-based inputs reach 96.5--97.9\% while direct LLM inputs reach 94.4--95.8\%, a 1.4--2.1 pp sketch margin. Both LLM-generated input strategies outperform random fuzzing (88.7--89.4\%) and example-only inputs (85.9--86.6\%) by roughly 6--11 pp. The main split is therefore not between sketching and direct concrete generation, but between LLM-generated diverse inputs and random fuzzing or the few examples in the prompt. The sketch abstraction still gives the strongest result in every column, suggesting that explicitly enumerating behavioral categories provides a small, consistent robustness gain over direct concrete generation.

Within each strategy, the three execution-based aggregators agree to within 1.4 pp. The strategy$\times$method interaction is therefore small relative to the input-source effect: the input source determines most of the behavioral-signal quality independently of how that signal is aggregated.

\subsection{Aggregation rules: HumanEval+}
\label{sec:main-results}

We now fix the input source to sketch-based generation and compare aggregation rules across the full $3\times 3$ grid (Table~\ref{tab:main-results}).

\begin{table}[H]
\begin{center}
\resizebox{\columnwidth}{!}{%
\begin{tabular}{ll cccccc}
\toprule
\textbf{Model} & \textbf{Think} & \textbf{Greedy} & \textbf{Best-of-$N$} & \textbf{Maj.\ Vote} & \textbf{Wt.\ Vote} & \textbf{MBR-Exec} & \textbf{\methodsc{}} \\
\midrule
\multirow{3}{*}{3.1 Pro}
& Low    & \multirow{3}{*}{93.9} & \multirow{3}{*}{97.0} & 75.0 & \textbf{98.8} & 97.6 & \textbf{98.8} \\
& Medium &      &      & 77.4 & \textbf{98.2} & 96.3 & 97.6 \\
& High   &      &      & 76.2 & \textbf{97.6} & 95.7 & 97.0 \\
\midrule
\multirow{3}{*}{3 Flash}
& Low    & \multirow{3}{*}{95.7} & \multirow{3}{*}{\textbf{96.3}} & 64.6 & \textbf{96.3} & 95.1 & 95.7 \\
& Medium &      &      & 65.9 & \textbf{96.3} & 95.1 & \textbf{96.3} \\
& High   &      &      & 76.2 & 94.5 & 93.9 & 95.1 \\
\midrule
\multirow{3}{*}{3.1 Flash Lite}
& Low    & \multirow{3}{*}{75.6} & \multirow{3}{*}{90.9} & 51.8 & \textbf{92.7} & \textbf{92.7} & 92.1 \\
& Medium &      &      & 43.9 & \textbf{93.3} & 92.7 & 92.7 \\
& High   &      &      & 52.4 & \textbf{93.3} & 92.7 & 92.7 \\
\bottomrule
\end{tabular}%
}
\end{center}
\caption{Pass@1 (\%) on HumanEval+ with $N=50$ candidates. The three execution-based aggregators (WV, MBR-Exec, \methodsc{}) agree to within 1--2 pp on every configuration and are statistically indistinguishable (Appendix~\ref{app:bootstrap}); all three exceed output-pattern MV by 18--49 pp. Greedy and Best-of-$N$ are reported as per-model baselines on this benchmark (one cached run per model reused across thinking levels; see Section~\ref{sec:experiments}).}
\label{tab:main-results}
\end{table}

\paragraph{Execution-based versus all-or-nothing voting.} Output-pattern MV scores 43.9--77.4\% across configurations, consistently below Best-of-$N$ (90.9--97.0\%); even greedy decoding on Pro (93.9\%) exceeds MV on every Pro configuration. The gap to the best execution-based method ranges from 19 pp (Flash high vs.\ \methodsc{}) to 49 pp (Lite medium vs.\ WV) and is largest on weaker models, where more generated candidates hit at least one generated-input error. The main failure mode is the all-success filter: once every candidate has at least one error on the generated inputs, MV returns no candidate, whereas exception-aware execution methods can still use partial agreement.

\paragraph{Effect of AST normalization.} One alternative explanation is that exact output-pattern voting is too sensitive to representation and that a stronger non-execution normalization would close the gap. We test this with \emph{AST-normalized MV}, which parses each candidate, alpha-renames variables, strips docstrings, and groups by canonical AST structure. On three representative configurations (Pro low, Flash medium, Lite low), AST-MV recovers most of the gap to output-pattern MV but still trails the best execution-based aggregator by 1--3 pp (Table~\ref{tab:ast-mv}). AST normalization rescues naming and formatting differences but cannot unify structurally different implementations that compute the same behavior, so its residual deficit is consistent with the need for behavioral evidence.

\begin{table}[H]
\begin{center}
\small
\begin{tabular}{l ccc}
\toprule
\textbf{Method} & \textbf{Pro low} & \textbf{Flash medium} & \textbf{Lite low} \\
\midrule
Output-pattern MV   & 75.0 & 65.9 & 51.8 \\
AST-normalized MV   & 95.7 & 94.5 & 90.9 \\
Best exec.-based    & \textbf{98.8} & \textbf{96.3} & \textbf{92.7} \\
\quad ($\Delta$ AST $-$ best exec.) & $-$3.1 & $-$1.8 & $-$1.8 \\
\bottomrule
\end{tabular}
\end{center}
\caption{Pass@1 (\%) on HumanEval+ for output-pattern MV, AST-normalized MV, and the best execution-based aggregator (WV or \methodsc{}). AST normalization closes 20--40 pp of the gap to execution-based methods but leaves a residual 2--3 pp deficit because it cannot unify structurally different but behaviorally equivalent programs.}
\label{tab:ast-mv}
\end{table}

\paragraph{Aggregation rules within the execution-based family.} The three execution-based aggregators are within 1--2 pp on every configuration. Bootstrap analysis (10{,}000 problem-level resamples; Appendix~\ref{app:bootstrap}) shows the SV--WV difference is in $[-0.79, +0.61]$ pp and not significant on any of the 18 configurations ($p > 0.05$); the SV--MBR-Exec difference is in $[-0.61, +1.22]$ pp with the same conclusion. The convergence is consistent with the oracle-gap analysis (Section~\ref{sec:oracle-gap}): with selection-failure rates already at 1.5--2.7\%, the residual room for any aggregator to differentiate is small.

\paragraph{Thinking level.} On HumanEval+, output-pattern MV benefits from deeper thinking in the clearest Flash case (64.6\% $\to$ 76.2\%) and peaks at medium on Pro (75.0\% $\to$ 77.4\%). Execution-based methods stay flat or decrease slightly: on Pro, both WV and \methodsc{} drop from 98.8\% (low) to 97.6\%/97.0\% (high); on Flash, WV drops from 96.3\% (low) to 94.5\% (high). Deeper thinking appears to reduce candidate diversity, leaving fewer distinct behavioral clusters to exploit. This is an interaction effect rather than a universal monotonic gain, as the MBPP+ results below show.

\subsection{Aggregation rules: MBPP+}
\label{sec:mbpp-results}

\begin{table}[t]
\begin{center}
\resizebox{\columnwidth}{!}{%
\begin{tabular}{ll cccccc}
\toprule
\textbf{Model} & \textbf{Think} & \textbf{Greedy} & \textbf{Best-of-$N$} & \textbf{Maj.\ Vote} & \textbf{Wt.\ Vote} & \textbf{MBR-Exec} & \textbf{\methodsc{}} \\
\midrule
\multirow{3}{*}{3.1 Pro}
& Low    & \multirow{3}{*}{96.6} & 96.8 & 77.2 & 96.8 & 96.8 & \textbf{97.4} \\
& Medium &      & 95.8 & 75.4 & 95.5 & 95.5 & 95.8 \\
& High   &      & 95.2 & 74.6 & 95.2 & 95.0 & 95.5 \\
\midrule
\multirow{3}{*}{3 Flash}
& Low    & \multirow{3}{*}{95.2} & \textbf{95.5} & 59.5 & 95.0 & 95.0 & 95.2 \\
& Medium &      & 92.6 & 51.3 & 92.3 & 92.3 & 91.8 \\
& High   &      & 92.9 & 52.9 & 91.5 & 91.3 & 91.0 \\
\midrule
\multirow{3}{*}{3.1 Flash Lite}
& Low    & \multirow{3}{*}{91.3} & 91.0 & 39.4 & \textbf{91.5} & 91.3 & 90.7 \\
& Medium &      & 91.8 & 41.0 & \textbf{92.1} & 91.8 & 91.3 \\
& High   &      & 91.5 & 42.6 & \textbf{92.1} & 91.3 & 91.5 \\
\bottomrule
\end{tabular}%
}
\end{center}
\caption{Pass@1 (\%) on MBPP+ (375--378 problems per config, $N=50$). The three execution-based aggregators are within 1--2 pp on every configuration and statistically indistinguishable (Appendix~\ref{app:bootstrap}); output-pattern MV trails by 19--52 pp. Greedy is reported as a per-model baseline (one cached run per model reused across thinking levels; Section~\ref{sec:experiments}). On MBPP+ the Best-of-$N$ pool is resampled per thinking level, so the Best-of-$N$ column does vary.}
\label{tab:mbpp-results}
\end{table}

The HumanEval+ patterns replicate: output-pattern MV scores 39.4--77.2\%, the three execution-based aggregators stay within 1--2 pp of each other, and the gap is largest on Flash Lite (39--43\% vs.\ 91--92\%). On MBPP+, SV is slightly ahead on Pro (+0.27 to +0.53 pp) and WV is slightly ahead on Flash Lite ($-$0.53 to $-$0.79 pp), but bootstrap analysis finds no significant difference on any configuration. A candidate-pool effect is also visible: on Flash medium and high, greedy decoding (95.2\%) exceeds the execution-based selection methods (91.0--92.3\%), and the same pattern appears more mildly on Pro medium/high. Temperature-0.8 sampling introduces diversity at the cost of per-candidate quality, and selection cannot recover quality absent from the pool.

\subsection{Oracle-gap decomposition}
\label{sec:oracle-gap}

We decompose failures into \emph{generation failures} (no correct candidate in the pool) and \emph{selection failures} (a correct candidate exists but is not chosen).

\begin{table}[t]
\begin{center}
\begin{tabular}{l ccc ccc}
\toprule
& \multicolumn{3}{c}{\textbf{Pass@1 (\%)}} & \multicolumn{3}{c}{\textbf{Failure rate (\%)}} \\
\cmidrule(lr){2-4} \cmidrule(lr){5-7}
\textbf{Config} & Oracle & SV & WV & Gen.\ & SV sel.\ & WV sel.\ \\
\midrule
Pro low      & 98.8 & 98.8 & 98.8 & 1.2 & 0.0 & 0.0 \\
Flash med    & 98.8 & 96.3 & 96.3 & 1.2 & 2.4 & 2.4 \\
Lite low     & 93.9 & 92.1 & 92.7 & 6.1 & 1.8 & 1.2 \\
\midrule
\multicolumn{4}{l}{\textit{Avg.\ across 9 configs (HumanEval+)}} & 2.8 & 1.8 & 1.5 \\
\multicolumn{4}{l}{\textit{Avg.\ across 9 configs (MBPP+)}} & 3.9 & 2.7 & 2.5 \\
\bottomrule
\end{tabular}
\end{center}
\caption{Oracle-gap decomposition. Generation failures dominate; selection failures average 1.5--2.7\% across all configurations. On Pro low, both SV and WV match the oracle ceiling.}
\label{tab:oracle-gap}
\end{table}

Selection failures average 1.5--2.7\% across the 18 configurations; generation failures account for the rest. Generation is therefore the larger bottleneck, which helps explain the convergence among execution-based aggregators: with $\le$3\% of problems available for an aggregator-specific improvement, observable differences among aggregation rules are limited.

\paragraph{Cluster diagnostics.} \methodsc{} forms 1.2--1.5 clusters per problem on average, with the largest cluster containing 38--48 of the 50 candidates. Across the 9 HumanEval+ configurations, execution-based methods solve 31--81 more problems than output-pattern MV, and MV solves zero problems that all three execution-based methods miss.

\paragraph{Selection failures are systematic.} Across the 18 configurations, \methodsc{} makes 119 selection errors on 52 unique problems; 34 problems recur in $\ge 2$ configurations. Failed problems have an average largest-cluster size of 25.6 (versus 38--48 in the typical case), and greedy decoding rescues 88 of the 119 errors. The correct solution exists but is not the plurality behavior in the sampled pool---a limitation common to all consensus-based selection.

\paragraph{Oracle computation in ablations.} In Table~\ref{tab:ablation-strategy}, the oracle is computed on the subset of problems where input generation succeeded for all four strategies. Because this subset excludes problems where sketch inputs failed, accuracy on the full benchmark can occasionally exceed the subset oracle---an artifact of the restricted denominator, not a logical inconsistency.

\subsection{\texorpdfstring{$D$}{D}-scaling}
\label{sec:ablation-d-scaling}

In a separate Flash-medium sweep with $N{=}50$ and sketch inputs, we vary $D \in \{5, 10, 20, 30, 50\}$ on HumanEval+ (Table~\ref{tab:d-scaling}). Pass@1 rises from 95.1\% at $D{=}5$ to 96.3\% at $D{=}10$ and saturates at 97.0\% from $D{=}30$ onward. Even five well-chosen inputs already exceed output-pattern MV by 29 pp on Flash medium (95.1\% vs.\ 65.9\%). Beyond $D{=}20$, the marginal gain is negligible: the average number of behavioral clusters per problem grows only from 1.20 ($D{=}5$) to 1.43 ($D{=}50$), confirming that on this benchmark a small number of well-chosen inputs is enough to separate the dominant correct cluster from buggy variants. We use $D{=}50$ as a conservative default since execution is cheap relative to candidate generation.

\begin{table}[t]
\begin{center}
\small
\begin{tabular}{l ccccc}
\toprule
\textbf{$D$} & 5 & 10 & 20 & 30 & 50 \\
\midrule
Pass@1 (\%)            & 95.1 & 96.3 & 96.3 & 97.0 & 97.0 \\
Avg.\ clusters/problem & 1.20 & 1.28 & 1.34 & 1.38 & 1.43 \\
Avg.\ largest cluster  & 48.5 & 47.8 & 47.6 & 47.4 & 47.2 \\
\bottomrule
\end{tabular}
\end{center}
\caption{$D$-scaling on HumanEval+ (Flash medium, $N{=}50$, sketch inputs). Pass@1 saturates at $D{\ge}30$; cluster diagnostics show inputs continue to fragment behavior past saturation, but the marginal new clusters do not contain new correct programs.}
\label{tab:d-scaling}
\end{table}

\subsection{Qualitative observations}
\label{sec:qualitative}

Two patterns recur when \methodsc{} differs from output-pattern MV. \emph{Partial behavioral evidence is preserved}: if every candidate raises on at least one generated input, output-pattern MV has no all-success survivor, while \methodsc{} can still select the plurality exception-aware fingerprint. \emph{Subtle bugs are separated}: two \texttt{divisors(n)} candidates differing only by \texttt{range(1, n+1)} vs.\ \texttt{range(1, n)} agree on $n \in \{1,2,3\}$, but the sketch-generated input $n=6$ separates the off-by-one variant.

\section{Discussion}
\label{sec:discussion}

\paragraph{Contribution scope.} Fingerprint clustering, weighted voting, and MBR-Exec all predate this work. Our contribution is the decomposition: input quality and aggregation rule are separate factors, and on these benchmarks input quality has the larger effect. The same input-quality gain transfers across SV, WV, and MBR-Exec, while no aggregator compensates for low-quality inputs. The convergence among SV/WV/MBR is itself part of the result: with selection-failure rates at 1.5--2.7\%, room for any aggregator to differentiate is narrow, and this convergence is conditional on a shared, discriminative execution trace.

\paragraph{Interpreting direct input generation.} Direct LLM input generation is close to sketch generation in our runs, indicating that a capable model can often produce useful behavioral probes without an explicit sketch stage. The remaining sketch margin is consistent across all nine cells of Table~\ref{tab:ablation-strategy}. The narrower conclusion: sketching is a reliable way to make behavioral coverage explicit, but the larger effect is the move from sparse examples or random values to LLM-generated diverse inputs.

\paragraph{Practical guidance and overhead.} The ablations suggest a simple ordering: replace prompt examples with generated probes first; prefer structured generation when the input domain has clear equivalence classes; increase $D$ only after checking that new inputs create new useful clusters (most of the gain appears by $D{=}10$). Fingerprinting executes $N{\times}D{=}2{,}500$ sandboxed programs per problem in 10--30\,s of local time versus 90--700\,s of LLM time; sketch-based input generation costs one additional LLM call.

\section{Related work}
\label{sec:related}

\paragraph{Self-consistency and majority voting.}
\citet{wang2023selfconsistency} introduced self-consistency for chain-of-thought reasoning; a common code adaptation clusters candidates by exact output-pattern equality after execution \citep{chen2021codex}. \citet{wang2025rankedvoting} aggregate ranked answer lists, but remain answer-text based and target reasoning rather than code.

\paragraph{Execution-based code selection.}
CodeT \citep{chen2023codet} jointly generates tests and code; MBR-Exec \citep{shi2022mbr} applies minimum Bayes risk on provided examples; \citet{jiang2026cascade} demonstrate LLM effectiveness in code understanding and transformation tasks, using LLMs to guide deterministic code transformations; AlphaCode \citep{li2022alphacode} clusters outputs on example tests and scales to $10^6$ candidates. We include MBR-Exec and compare against example-only inputs (Table~\ref{tab:ablation-strategy}). \methodsc{} differs by using sketch-based LLM-generated inputs and exact-equality fingerprint clustering.

\paragraph{Test input generation.}
Fuzzing \citep{miller1990fuzz} and property-based testing \citep{claessen2000quickcheck} generate inputs automatically. LLM-based test generation has been studied by \citet{chen2023codet}, \citet{hong2025alloy} and \citet{jiang2024javadoc}. Our sketch-based approach structures generation around behavioral categories rather than direct value generation.

\paragraph{Program equivalence and inference-time scaling.}
Differential testing \citep{mckeeman1998differential}, program sketching \citep{solar2013program}, and inference-time scaling for code \citep{snell2025scaling} are related reference points; OBsmith \citep{jiang2026obsmith} and APRIL \citep{zhong2025apisynth,zhong2025april} apply LLM-driven sketching and prompt optimization in adjacent settings. \methodsc{} improves the selection step within a fixed sampling budget.

\section{Conclusion}
\label{sec:conclusion}

Across three Gemini models, three thinking levels, and two benchmarks, inference-time code selection behaves primarily as a signal-quality problem. The best execution-based selector outperforms output-pattern majority voting by 19--52 pp on every configuration, and output-pattern MV consistently underperforms Best-of-$N$. Sketch-based input generation is consistently strongest in the input-strategy ablation, with a 0.6--2.1 pp margin over direct LLM input generation and up to 11.3 pp over random fuzzing or example-only inputs. These input-source effects transfer across \methodsc{}, weighted voting, and MBR-Exec, which are statistically indistinguishable ($p > 0.05$) once inputs are fixed.

Operationally, oracle-free selection benefits from exception-aware behavioral traces, input quality matters more than the execution-based aggregator, low-to-moderate thinking levels often work best for execution-based selection, and $D \approx 10$ well-chosen inputs captures most of the gain in our sweep. \methodsc{} is the instrument used to evaluate these claims, but the main contribution is the decomposition.

\paragraph{Limitations.} We evaluate Gemini models on two Python benchmarks; other model families or languages require new runs and, for non-Python tasks, a new sandbox. Sketch-based input generation depends on the LLM's domain familiarity, so specialized APIs may reduce the input-strategy advantage. Programs with intentional nondeterminism produce inconsistent fingerprints; we do not evaluate sandbox seeding as a mitigation.


\bibliographystyle{plainnat}
\bibliography{references}

\appendix
\section{Prompts}
\label{app:prompts}

We provide the exact prompts used in our pipeline. All prompts are zero-shot with no in-context examples beyond the format specification.

\subsection{Candidate generation prompt}

Used to generate $N$ candidate solutions for each problem. The \texttt{\{prompt\}} placeholder is replaced with the benchmark problem's function signature and docstring.

\begin{verbatim}
Complete the following Python function.
Write ONLY the function body (the lines
that go inside the function). Do NOT
repeat the function signature, docstring,
or imports. Do NOT use markdown fences.

{prompt}
\end{verbatim}

\subsection{Sketch input generation prompt}

Used to generate $K$ diverse input sketches (abstract input categories) for a given function. The \texttt{\{K\}} placeholder is replaced with the desired number of sketches.

\begin{verbatim}
You are generating diverse test inputs
for a Python function.

Function signature and description:
{problem_description}

Generate {K} diverse INPUT SKETCHES. Each
sketch should target a fundamentally
different equivalence class of behavior.
Think about:
- Edge cases (empty, single element,
  None, zero, negative)
- Boundary values (max int, very long
  strings, deeply nested)
- Typical cases (medium-sized, mixed types)
- Special structure (sorted, reversed,
  all duplicates, alternating)

For each sketch, provide:
1. A short description of what case it tests
2. A concrete Python expression that
   produces a valid input

Output as JSON array:
[{"description": "...", "input_expr": "..."}]

Generate EXACTLY {K} sketches.
\end{verbatim}

\subsection{Input variation prompt}

Used to instantiate $M-1$ additional concrete inputs from each sketch.

\begin{verbatim}
Given this input sketch for a Python
function:
Description: {description}
Example: {input_expr}

Function info:
{problem_description}

Generate {M} more concrete inputs that
follow the SAME pattern but with different
specific values. Output as a JSON array
of Python expressions.

Return ONLY the JSON array, no other text.
\end{verbatim}

\section{Detailed results}
\label{app:full-results}

Across all 9 configurations on HumanEval+, execution-based methods (WV, MBR-Exec, \methodsc{}) solve 31--81 more problems than output-pattern majority voting, while MV never solves a problem that any execution-based method misses. The oracle upper bound (percentage of problems where at least one candidate is correct) ranges from $\sim$93.9\% (Flash Lite) to $\sim$99\% (Pro), confirming that the remaining errors for the best methods are generation failures, not selection failures.

Results on MBPP+ (Table~\ref{tab:mbpp-results}) replicate the HumanEval+ patterns on a larger benchmark; SV and WV trade small leads across configurations and are never statistically distinguishable. The input-generation-strategy ablation (Table~\ref{tab:ablation-strategy}) and the $D$-scaling sweep (Section~\ref{sec:ablation-d-scaling}) together identify input quality as the critical component, with LLM-generated diverse inputs outperforming weaker input sources and input quality dominating both input quantity and the choice of aggregation rule.

\paragraph{API usage.}
Each problem uses approximately 10K input tokens and 10K output tokens for $N=50$ candidate-generation calls, plus approximately 500 input and 500 output tokens for sketch input generation. Dollar costs should be recomputed from the provider's current pricing for the exact preview endpoint used; preview model availability and pricing can change. Execution and fingerprinting are local and do not require additional API calls.

\section{Bootstrap confidence intervals}
\label{app:bootstrap}

To rigorously assess whether \methodsc{} differs from the other execution-based aggregators in performance, we compute bootstrap confidence intervals using 10{,}000 problem-level resamples with replacement. For each resample, we compute pass@1 for the methods of interest and record the difference. Tables~\ref{tab:bootstrap-he} and~\ref{tab:bootstrap-mbpp} report SV vs.\ WV on HumanEval+ and MBPP+ respectively; the SV vs.\ MBR-Exec comparison is summarized in the closing paragraph of this appendix.

\begin{table}[H]
\begin{center}
\resizebox{\columnwidth}{!}{%
\begin{tabular}{ll cc c cc}
\toprule
\textbf{Model} & \textbf{Think} & \textbf{SV} & \textbf{WV} & \textbf{$\Delta$ (SV$-$WV)} & \textbf{95\% CI} & \textbf{$p$-value} \\
\midrule
\multirow{3}{*}{3.1 Pro}
& Low    & 98.78 & 98.78 & $+$0.00 & [$-$0.00, $+$0.00] & 1.000 \\
& Medium & 97.56 & 98.17 & $-$0.61 & [$-$1.83, $+$0.00] & 0.370 \\
& High   & 96.95 & 97.56 & $-$0.61 & [$-$1.83, $+$0.00] & 0.376 \\
\midrule
\multirow{3}{*}{3 Flash}
& Low    & 95.73 & 96.34 & $-$0.61 & [$-$3.05, $+$1.22] & 0.394 \\
& Medium & 96.34 & 96.34 & $+$0.00 & [$-$1.83, $+$1.83] & 0.655 \\
& High   & 95.12 & 94.51 & $+$0.61 & [$+$0.00, $+$1.83] & 0.361 \\
\midrule
\multirow{3}{*}{3.1 Flash Lite}
& Low    & 92.07 & 92.68 & $-$0.61 & [$-$1.83, $+$0.00] & 0.358 \\
& Medium & 92.68 & 93.29 & $-$0.61 & [$-$1.83, $+$0.00] & 0.358 \\
& High   & 92.68 & 93.29 & $-$0.61 & [$-$1.83, $+$0.00] & 0.358 \\
\bottomrule
\end{tabular}%
}
\end{center}
\caption{Bootstrap analysis on HumanEval+ (164 problems, 10{,}000 resamples). No SV--WV difference is statistically significant at $\alpha = 0.05$. All differences fall within $\pm$0.61 percentage points.}
\label{tab:bootstrap-he}
\end{table}

\begin{table}[H]
\begin{center}
\resizebox{\columnwidth}{!}{%
\begin{tabular}{ll cc c cc}
\toprule
\textbf{Model} & \textbf{Think} & \textbf{SV} & \textbf{WV} & \textbf{$\Delta$ (SV$-$WV)} & \textbf{95\% CI} & \textbf{$p$-value} \\
\midrule
\multirow{3}{*}{3.1 Pro}
& Low    & 97.34 & 96.81 & $+$0.53 & [$+$0.00, $+$1.33] & 0.133 \\
& Medium & 95.74 & 95.48 & $+$0.27 & [$-$0.53, $+$1.06] & 0.388 \\
& High   & 95.47 & 95.20 & $+$0.27 & [$+$0.00, $+$0.80] & 0.368 \\
\midrule
\multirow{3}{*}{3 Flash}
& Low    & 95.24 & 94.97 & $+$0.26 & [$-$0.53, $+$1.32] & 0.394 \\
& Medium & 91.80 & 92.33 & $-$0.53 & [$-$1.32, $+$0.00] & 0.137 \\
& High   & 91.01 & 91.53 & $-$0.53 & [$-$1.32, $+$0.00] & 0.135 \\
\midrule
\multirow{3}{*}{3.1 Flash Lite}
& Low    & 90.74 & 91.53 & $-$0.79 & [$-$2.12, $+$0.26] & 0.118 \\
& Medium & 91.27 & 92.06 & $-$0.79 & [$-$1.85, $+$0.00] & 0.051 \\
& High   & 91.53 & 92.06 & $-$0.53 & [$-$1.59, $+$0.53] & 0.233 \\
\bottomrule
\end{tabular}%
}
\end{center}
\caption{Bootstrap analysis on MBPP+ (375--378 problems per config, 10{,}000 resamples). No SV--WV difference is statistically significant at $\alpha = 0.05$, though Flash Lite medium ($p = 0.051$) is borderline. SV is consistently ahead of WV on Pro ($+$0.27 to $+$0.53 pp) but behind on Flash Lite ($-$0.53 to $-$0.79 pp).}
\label{tab:bootstrap-mbpp}
\end{table}

Across all 18 configurations, the SV--WV difference never exceeds $\pm$0.79 percentage points and is never statistically significant at $\alpha = 0.05$. The same bootstrap procedure applied to SV vs.\ MBR-Exec yields differences in $[-0.61, +1.22]$ pp, also non-significant on all 18 configurations ($p > 0.05$). This confirms that \methodsc{}, weighted voting, and MBR-Exec occupy the same statistical tier. By contrast, the gap between any execution-based method and output-pattern majority voting (18--52 points) is far larger, reinforcing our central finding that the choice of execution-based \emph{vs.}\ output-pattern selection is more consequential than the choice \emph{among} execution-based aggregation rules, as predicted by the signal-quality framing.

\end{document}